\documentclass[review, aps ,amsmath, amssymb]{elsarticle}

\usepackage{hyperref}

\usepackage{graphicx}
\usepackage{dcolumn}
\usepackage{bm}
\usepackage{color}

\begin{document}

\begin{frontmatter}

\title{Parametric Lattice Boltzmann Method}

\author{Jae Wan Shim\corref{mycorrespondingauthor}}
\address{Materials and Life Science Research Division, Korea Institute of Science and Technology; and Major of Nanomaterials Science and Engineering, KIST Campus, Korea University of Science and Technology, 5 Hwarang-ro 14-gil Seongbuk Seoul 02792 Republic of Korea}
\cortext[mycorrespondingauthor]{Corresponding author}

\begin{abstract}
The discretized equilibrium distributions of the lattice Boltzmann method are presented by using the coefficients of the Lagrange interpolating polynomials that pass through the points related to discrete velocities and using moments of the Maxwell-Boltzmann distribution. The ranges of flow velocity and temperature providing positive valued distributions vary with regulating discrete velocities as parameters. New isothermal and thermal compressible models are proposed for flows of the level of the isothermal and thermal compressible Navier-Stokes equations. Thermal compressible shock tube flows are simulated by only five on-lattice discrete velocities. Two-dimensional isothermal and thermal vortices provoked by the Kelvin-Helmholtz instability are simulated by the parametric models.
\end{abstract}


\end{frontmatter}

\section{Introduction}
One way of simulating fluid flows is to use artificial particles jumping from one node to another in a regular lattice with a limited number $q$ of discrete velocities as in the lattice Boltzmann method \cite{FHP,McNamara,Higuera,Chen,Chen1992,Qian1992}. At a given node $x$ and time $t$, the existence of a particle having a given discrete velocity $v_i$  is expressed by a probability $p_i(x,t)$ in real numbers instead of zero or one. Hence, the density of the particles having $v_i$ is
\begin{equation}
\label{eq:density}
f_i(x,t)=\rho(x,t) p_i(x,t)
\end{equation}
where $\rho(x,t)$ is a total density. Particles collide with each other every time step $\Delta t$ and thus velocity distributions change according to a given redistribution rule $r_i(x,t)$ or a discretized equilibrium distribution,
\begin{equation}
\label{eq:eqdist}
f^{eq}_i(x,t)=\rho(x,t) r_i(x,t),
\end{equation}
within the following discretized advection formula having a single relaxation constant $\omega$ as
\begin{equation}
\label{eq:srt}
f_i(x+v_i\Delta t,t+\Delta t)=(1-\omega)f_i(x,t)+\omega f^{eq}_i(x,t).
\end{equation}
The constitution of $f^{eq}_i(x,t)$ with corresponding discrete velocities $v_i$ affects the accuracy, efficiency, and stability of the lattice Boltzmann method. We will present a new general form of $f^{eq}_i$ for the purpose of simulating flows of the level of the Navier-Stokes equations
\begin{equation} 
\label{eq:ns}
\left\{ 
\begin{array}{l}
\partial_t\rho+\nabla\cdot(\rho u)=0, \\
\partial_t(\rho u)+\nabla\cdot(\rho u \otimes u)=\nabla \cdot (\mathbf S-\rho \theta \mathbf I),\\
\partial_t\left(\frac d 2 \rho \theta\right)+\nabla\cdot\left(\frac d 2\rho \theta u\right)+\nabla \cdot \mathbf q=\mathbf S: (\nabla u)-\rho \theta \nabla \cdot u
\end{array} \right.
\end{equation}
with
\begin{eqnarray}
\mathbf S&=& \nu\rho(\nabla u+\nabla u^\mathsf T-\frac 2 d \nabla \cdot u \mathbf I)+\eta\nabla \cdot u \mathbf I, \nonumber \\
\mathbf q&=&-\kappa \nabla \theta \nonumber 
\end{eqnarray}
where $\theta \equiv kT/m$ with $k$ being the Boltzmann constant, $T$ the Kelvin temperature, and $m$ mass of a particle, $u$ is flow velocity, $d$ dimension of space, $\nu$ kinematic viscosity, $\eta$ bulk viscosity, and $\kappa$ thermal conductivity. The general form is not limited to provide models up to this level but beyond by increasing the number of discrete velocities $q$. 

\section{Parametric discretized equilibrium distribution}
\subsection{General form}
Here, we present new discretized equilibrium distributions $f^{eq}_i$, namely parametric discretized equilibrium distributions. For simplicity, we present $r_i$ that gives $f^{eq}_i$ in one-dimensional space according to Eq.~(\ref{eq:eqdist}) as
\begin{equation}
\label{eq:rdr}
r_i=\sum_{j=1}^q c_{ij} \mu_{j-1}
\end{equation}
where $c_{ij}$ is the coefficient corresponding to the term of degree $j-1$ of the Lagrange interpolating polynomial that passes through $(v_k,\delta_{ik})$ for $k=1,2,\ldots,q$ in which $\delta_{ik}$ is the Kronecker delta and $\mu_{n}$ is the $n$th moment of the Maxwell-Boltzmann distribution $F(v)$ defined by $\mu_n=\int v^{n}F(v) dv$. By defining $\hat \mu_n=\sum v_i^n r_i$, this rule $r_i$ satisfies the $n$th moment identity $\hat \mu_n=\mu_n$ for $n=0,1,\ldots,q-1$ in one-dimensional space so that we have a relation between a desired order of accuracy $n^*$ and the number of discrete velocities $q$ as
\begin{equation}
n^*=q-1.
\end{equation}
 The detailed derivation is provided in Appendix. Multi-dimensional models can be obtained by tensor products of one-dimensional models or be directly derived from Eq.~(\ref{eq:constraints}) with proper choices of discrete velocities and a desired accuracy.

According to the Chapman-Enskog expansion \cite{chapman,chapmanDetail}, we obtain that a model satisfying $n^*=3$ recovers the isothermal compressible Navier-Stokes equations, namely the first two lines of Eq.~(\ref{eq:ns}) with bulk viscosity $\eta=0$ and kinematic viscosity
\begin{equation}
\nu=\left(\frac 1 \omega - \frac 1 2 \right)\theta\Delta t
\end{equation}
and a model satisfying $n^*=4$ recovers the thermal compressible Navier-Stokes equations, namely Eq.~(\ref{eq:ns}) with the same kinematic and bulk viscosities to the isothermal model and thermal conductivity
\begin{equation}
\kappa=\frac{d+2}{2}\nu\rho.
\end{equation}

\subsection{{Advantage of parametric models}}
The parametric lattice Boltzmann method(PLBM) provides a different way of deriving and a different point of view of understanding the existing models including the classic lattice Bhatnagar-Gross-Krook(LBGK) model \cite{Qian1992}. According to the framework provided by the PLBM, one can obtain, for a given number of discrete velocities, a set of lattice Boltzmann models which are equipped with parameters. For example, considering the models of three discrete velocities, one can obtain the LBGK model by fixing the parameter $\zeta=3$ in Eq.~(\ref{eq:1d3v}).

The new several models provided by the PLBM have advantages with respect to the existing counterpart models as the followings. One can obtain a new model with three discrete velocities, which is called the parametric model with $\zeta=4$ in this article. This model is more stable than the LBGK model and is more accurate than the entropic model. The formula, analysis, and benchmark test are described in the following sections and especially in Eq.~(\ref{eq:1d3v}), Table~\ref{tab:table3vel}, Figs.~\ref{fig:3v} to \ref{fig:LPConvergence}.

In addition, one can obtain a new model with four discrete velocities by the PLBM, which recovers the accuracy of the isothermal Navier-Stokes equations by the Chapman-Enskog expansion. Note that the three velocities models such as the LBGK and the entropic models do not recover the exact isothermal Navier-Stokes equations. The errors of these models are provided in Table~\ref{tab:tableAccuracy}. We also emphasize that the parametric four velocities models provide \emph{on-lattice} models in contrast to the existing off-lattice ones. Details are explained in the following subsection.

Moreover, the PLBM provides the thermal \emph{on-lattice} models which recover the accuracy of the thermal Navier-Stokes equations by \emph{only five} discrete velocities. We emphasize that the existing on-lattice models need seven discrete velocities and details are explained in the subsection containing Eq.~(\ref{eq:1d5v}). The benchmark tests are provided in the following sections and especially in Figs.~\ref{fig:PLBM5vshock} to \ref{fig:PLBM25v}.

\subsection{Example for isothermal flows}
As an example, $r_i$ of a model consisting of three discrete velocities $v_1=0$ and $v_{2,3}=\pm \sqrt {\zeta \theta_0}$ with a reference temperature $\theta_0$ can be expressed by 
\begin{equation}
\label{eq:1d3v}
r_i=w_i\left[1+\frac {v_i u}{\theta_0}+\frac {u^2}{(\zeta-1)\theta_0^2} (v_i^2-\theta_0)\right]
\end{equation}
with $w_1=1-1/\zeta$ and $w_{2,3}=1/(2\zeta)$ where $u$ is flow velocity distinguished from particle velocity $v$ and its discretized one $v_i$. Note two values of the parameter $\zeta=3$ and $4$ as in Table~\ref{tab:table3vel}. With the former, we recover the classical equilibrium distribution called the lattice Bhatnagar-Gross-Krook(LBGK) model \cite{Qian1992}, and with the latter, we find a more stable model in which the range of $u$ providing $r_i\geq 0$ is wider than any other value of $\zeta$. We will demonstrate its enhanced stability by a simulation of the shock tube and will discuss its accuracy.

With a set of four discrete velocities such as $v_{1,2}=\pm a$ and $v_{3,4}=\pm b$, we can obtain on-lattice models as
\begin{equation}
r_{i}=\frac{v_i c^2 \mu_0 + c^2 \mu_1- v_i \mu_2- \mu_3}{2 (v_i c^2-v_i^3)}
\end{equation}
where $c=b$ for $i=1$ and $2$ or $c=a$ for $i=3$ and $4$, which satisfy $n^*=3$ as in Table~\ref{tab:tableAccuracy} that is the condition to recover the accuracy of the isothermal Navier-Stokes equations by the Chapman-Enskog expansion. We can give $a=\sqrt {3\theta_0}/2$ when $b=3a$ to maximize the range of $u$ providing $r_i \geq 0$, for example.

Note that the three-velocities models including the LBGK do not recover the exact isothermal Navier-Stokes equations but have an error in viscous term -- the LBGK recovers the accuracy of the isothermal Navier-Stokes equations with the assumption of small $u$ to reduce the error of $u^3$ in Table~\ref{tab:tableAccuracy}.

We emphasize that the four-velocities parametric models provide \emph{on-lattice} models in contrast to the \emph{off-lattice} four-velocities model obtained by the conventional framework using the Gauss-Hermite quadrature. We will explain the concept of on- and off-lattice models in detail.

\begin{table*}
\caption{\label{tab:table3vel}The discretized equilibrium distributions $r_i$ of two specific models using three discrete velocities $v_1=0$ and $v_{2,3}=\pm \sqrt {\zeta \theta_0}$ are tabulated by using Eq.~(\ref{eq:1d3v}). Note that the parametric model with $\zeta=3$ is identical to the LBGK model. In this table, the LBGK and parametric models have symmetric discrete velocities. We can also obtain asymmetric models by using Eq.~(\ref{eq:rdr}). For example, when $v_1=0$, $v_1=2\sqrt {\theta_0}$, and $v_{3}=-4\sqrt {\theta_0}$, we have $r_1=\frac 1 8 [7-\frac{2u}{\sqrt \theta_0}-\frac {u^2}{\theta_0}]$, $r_2=\frac 1 {12}[1+ \frac{4u}{\sqrt \theta_0}+\frac{u^2}{\theta_0}]$, and $r_3=\frac 1 {24}[1-\frac{2u}{\sqrt\theta_0}+\frac{u^2}{\theta_0}]$. The accuracy of the models are provided in Table~\ref{tab:tableAccuracy}.}
\centering
\begin{tabular}{c|ccc}
\hline
Model&$r_1$&$r_2$&$r_3$\\ \hline 
LBGK($\zeta=3$)&$\frac{2}{3}[1-\frac{u^2}{2\theta_0}]$&$\frac{1}{6}[1+\frac{\sqrt{3}u}{\sqrt{\theta_0}}+\frac{u^2}{\theta_0}]$&$\frac{1}{6}[1-\frac{\sqrt{3}u}{\sqrt{\theta_0}}+\frac{u^2}{\theta_0}]$\\
Parametric($\zeta=4$)&$\frac{3}{4}[1-\frac{u^2}{3\theta_0}]$&$\frac{1}{8}[1+\frac{2u}{\sqrt{\theta_0}}+\frac{u^2}{\theta_0}]$&$\frac{1}{8}[1-\frac{2u}{\sqrt{\theta_0}}+\frac{u^2}{\theta_0}]$\\
\hline 
\end{tabular}
\end{table*}

\subsection{Example for thermal compressible flows}
As another example, thermal compressible flows of the Navier-Stokes equations can be simulated by only five on-lattice discrete velocities in one-dimensional space with the following rule and by 25 and 125 in two- and three-dimensional spaces via tensor products. For a symmetric set of discrete velocities defined by $v_1=0$, $v_{2,3}=\pm a$, and $v_{4,5}=\pm b$, the corresponding explicit expression of $r_i$ is
\begin{equation}
\label{eq:1d5v}
\left \{
\begin{array}{l}
r_1=\frac{\mu _0  a^2 b^2-\mu _2 (a^2+b^2)+\mu _4}{a^2 b^2},\\
r_{i \neq 1}=\frac{-\mu _1 v_i c^2-\mu _2 c^2+\mu
   _3 v_i+\mu _4}{2 v_i^2 (v_i^2-c^2)}
\end{array}
\right.
\end{equation}
where $c=b$ for $i=2$ and $3$ or $c=a$ for $4$ and $5$.
According to the Gauss-Hermite quadrature in the lattice Boltzmann theory \cite{Abe, He}, we can simulate thermal compressible flows with five discrete velocities obtained from the zeros $z_i$ of the Hermite polynomial of degree five \cite{Shan}, however, there is an important difference. While the ratios between $z_i (\neq 0)$ are not always rational so that artificial particles are not allowed to jump from one node to another in a regular lattice, the discrete velocities obeying the rule of Eq.~(\ref{eq:rdr}) are allowed to do so -- we call them on-lattice velocities -- by regulating $a$ and $b$ such as $b=2a$ in Eq.~(\ref{eq:1d5v}). For the on-lattice models, the conventional minimal sets consist of seven velocities for one-dimensional space \cite{Shim2013U}, and 37 velocities \cite{Philippi} or sparse 33 velocities \cite{Shim2013M,ShimZeit} for two-dimensional space in contrast to 25 velocities presented in this paper. 

\section{Analysis of the isothermal models}
\subsection{Ranges providing positive valued distributions}
\begin{figure}
\centering
\includegraphics{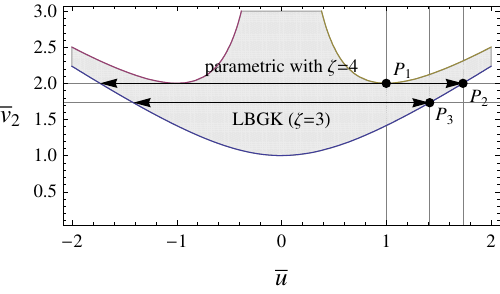}
\caption{\label{fig:3v} (Color online) The redistribution rule $r_i$ of three discrete velocities is drawn. The shadow area represents $r_i \geq 0$. The lower boundary passing through the points $P_2$ and $P_3$ represents $r_1=0$ and the two upper boundaries represent $r_{2,3}=0$. The point $P_1=(1,2)$ is touched by $\bar v_2=2$ or $\zeta=4$. The points $P_2=(\sqrt 3,2)$ and $P_3=(\sqrt 2,\sqrt 3)$ are the cross points of $\bar v_2=2$ (parametric model with $\zeta=4$) and $\bar v_2=\sqrt 3$ (LBGK model) with respect to $r_1=0$, respectively.}
\end{figure}

Let us define dimensionless variables $\bar u=u/\sqrt{\theta_0}$, $\bar v_i=v_i/\sqrt{\theta_0}$, and $\bar \theta=\theta/\theta_0$ for simplicity and examine Eq.~(\ref{eq:1d3v}). The contour plot of $r_i$ with respect to $\bar u$ and $\bar v_2 (=\sqrt \zeta)$ is shown in Fig.~\ref{fig:3v}. The shadow area represents the domains providing $r_i\geq0$. We observe that the range of $\bar u$ satisfying $r_i\geq 0$ for all $i$ is maximized as $| \bar u | \leq \sqrt 3$ when $\zeta=4$ or $\bar v_2=2$. Note that the range of the LBGK model is $| \bar u | \leq \sqrt 2$ and it is achieved when $\zeta=3$. 

\begin{figure}
\centering
\includegraphics{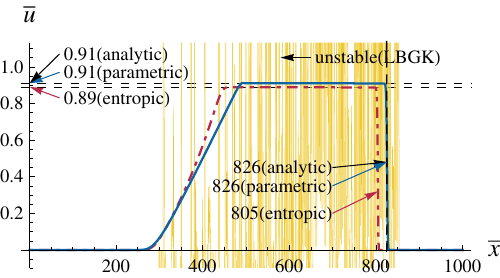}
\caption{\label{fig:PLE6} (Color online) The velocity profiles obtained by the LBGK model (yellow oscillating), the parametric model with $\zeta=4$ (blue solid line), and the entropic model (red dot-dashed) are drawn. The initial density of the left half space is $\bar \rho_L=6$ and that of the right is $\bar \rho_R=1$. For the whole space, the initial velocity and temperature are $\bar \theta_{L,R}=1$ and $\bar u_{L,R}=0$. The positions of the shock front $\bar x=$826 (analytic solution of Euler eq. \& parametric model with $\zeta=4$) and 805 (entropic model) and the post-shock velocity $\bar u=$0.91 (analytic \& parametric) and 0.89 (entropic) are indicated.}
\end{figure}

\begin{figure}
\centering
\includegraphics{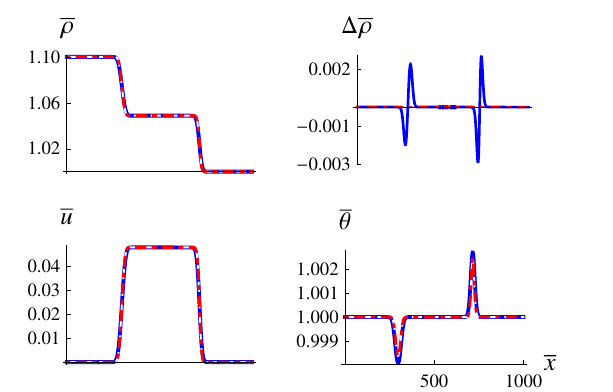}
\caption{\label{fig:PLEsmall} (Color online) The density $\bar \rho$, velocity $\bar u$, and temperature $\bar \theta$ profiles obtained by three discrete velocities with $\bar \rho_L/\bar \rho_R=1.1$ are drawn for the LBGK model (white dashed), the entropic model (red dot-dashed), and the parametric model with $\zeta=4$ (thick blue). The density difference $\Delta \bar \rho$ for the parametric model (thick blue) and for the entropic model (red dot-dashed) with respect to the LBGK model is provided for clarity. The maximum $\Delta \bar \rho$ of the parametric model with $\zeta=4$ with respect to the LBGK model is about $0.3$\%. Note that the horizontal axis label $\bar x$ is not always displayed for simplicity.}
\end{figure}

\begin{figure}
\centering
\includegraphics{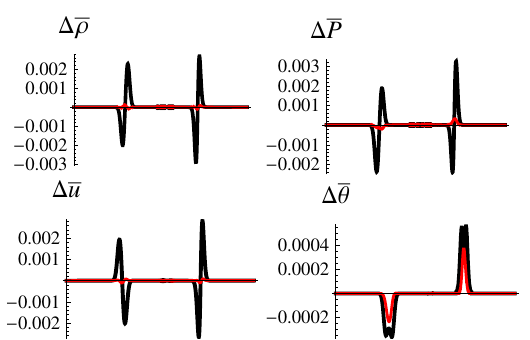}
\caption{\label{fig:PLBMviscosityMatch} (Color online) The differences of density $\Delta \bar \rho$, pressure $\Delta \bar P$, velocity $\Delta \bar u$, and temperature $\Delta \bar \theta$ profiles obtained by the parametric model with $\zeta=4$ and the LBGK model for the initial condition $\bar \rho_L/\bar \rho_R=1.1$ are drawn to demonstrate the enhancement of the viscosity matching by using $\hat \nu \equiv \nu(\zeta-1)/{2}$ instead of $\nu=(1/\omega-1/2)\sqrt {\theta_0}\Delta x/\sqrt{\zeta}$ by considering $\hat \mu_3=\zeta \theta_0 u$ of the parametric model. The thick black line corresponds to the difference between the LBGK model and the parametric model with $\zeta=4$ by using viscosity $\nu$. The thin red line is the result obtained by using $\hat \nu$ instead of $\nu$. We observe that the difference is significantly reduced in the case of using $\hat \nu$.}
\end{figure}

\begin{table*}
\caption{\label{tab:tableAccuracy}The moments of the Maxwell-Boltzmann distribution and of the LBGK, the parametric three-velocities with $\zeta=4$, the entropic, and the parametric four-velocities models for isothermal compressible flows, and of the parametric five-velocities model for thermal compressible flows are listed to compare accuracy of the models. Note that the recovery of the moments up to the 4th-order is the condition to recover the \emph{thermal} Navier-Stokes equations. Note that the temperature $\theta$ for the Maxwell-Boltzmann model is fixed to $\theta_0$ for the cases of isothermal models. Note that, as the footnote 1 of this table, the second-order moment of the entropic model could be expanded by the Taylor series expansion with respect to $u=0$ as $\theta_0+u^2-u^4/(4\theta_0)+\cdots$.}
\begin{tabular}{c|ccc}
\hline
Model&2nd-order&3rd-order&4th-order\\ \hline
Maxwell-Boltzmann&$\theta+u^2$&$3\theta u+u^3$&$3\theta^2+6 \theta u^2+u^4 $\\
LBGK($\zeta=3$)&$\theta_0+u^2$&$3\theta_0 u$&--\\
Parametric($\zeta=4$)&$\theta_0+u^2$&$4\theta_0 u$&--\\
Entropic&$-\theta_0+2\sqrt{\theta_0(\theta_0+u^2)}\footnote{It could be expanded by the Taylor series expansion with respect to $u=0$ as $\theta_0+u^2-u^4/(4\theta_0)+\cdots$.}$&$3\theta_0 u$&-- \\
Parametric 4-vel.&$\theta_0+u^2$&$3\theta_0 u+u^3$&--\\
Parametric 5-vel.&$\theta+u^2$&$3\theta u+u^3$&$3\theta^2+6 \theta u^2+u^4 $\\
\hline
 
\end{tabular}
\end{table*}

\begin{figure}
\centering
\includegraphics{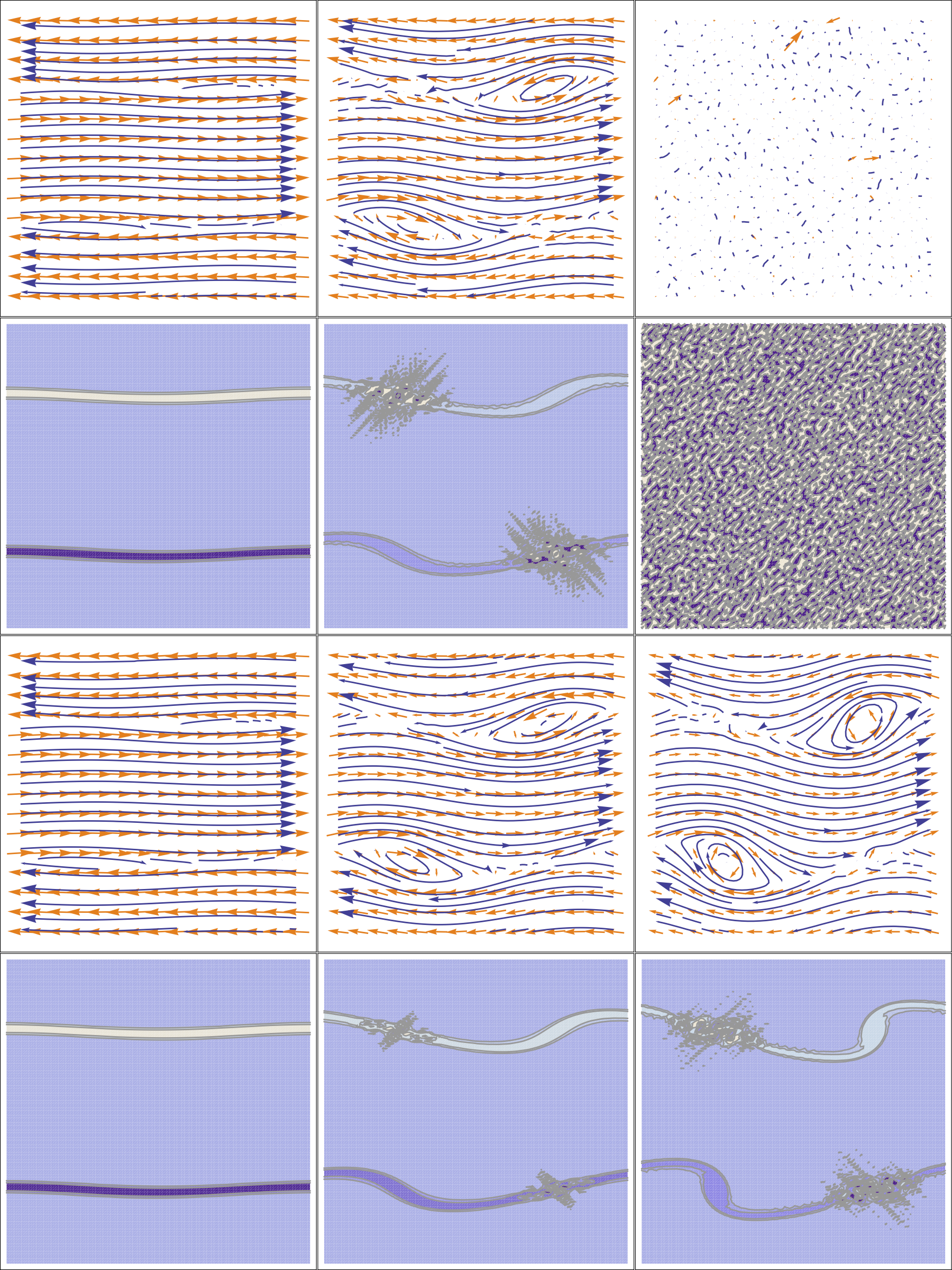}
\caption{\label{fig:LPViscosityMatch2d} (Color online) The shear layer simulation in two-dimensional space is presented by the tensor product of the parametric model with $\zeta=4$ and by the LBGK D2Q9 model. The shear layers provoke the Kelvin-Helmholtz instability so that vortices are generated. The first two and the last two rows are respectively obtained by the LBGK D2Q9 and the parametric models. The figures of the first and third rows show the velocity vectors (short orange arrows) with stream lines (long blue arrows) for time steps $500$, $1500$, and $1800$ (for the cases of the LBGK); and $577$, $1732$, and $2078$ (for the cases of the parametric model). The figures of the second and the fourth rows give the vorticity for the same time steps with the contours of $\pm(0.01,0.02,0.05)$. The result of the parametric model is slightly unstable at time step $1732$ (equivalent to 1500 for the LBGK) and we still observe vortices at time step $2078$ (equivalent to 1800 for the LBGK), however, that of the LBGK is already highly unstable at time step $1500$ and we only observe noise at time step $1800$.}
\end{figure}

\begin{figure}
\centering
\includegraphics{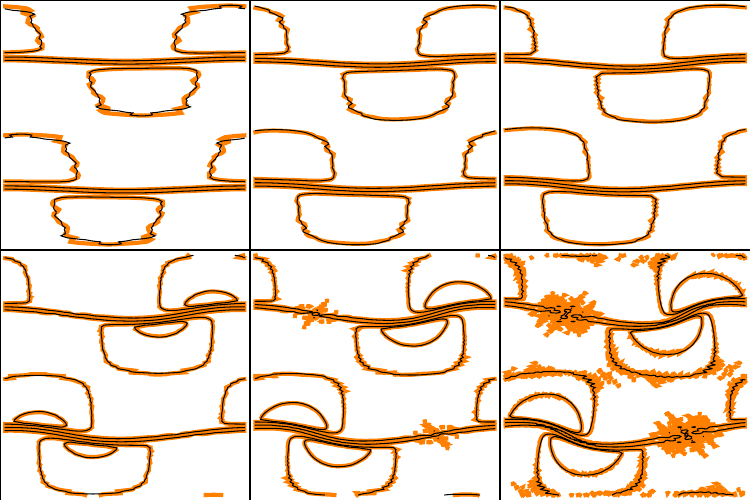}
\caption{\label{fig:lpCompare} (Color online) A comparison of the velocity amplitude results of the shear layer simulation in two-dimensional space obtained by the tensor product of the parametric model with $\zeta=4$ (black thin line) and by the LBGK D2Q9 model (orange thick line) is presented for the time steps from $500$ (left subfigure of the first row) to $1500$ (right subfigure of the second row) with intervals of $200$ by the LBGK step. The contours indicates the values of $0.03$, $0.07$, and $0.08$.}
\end{figure}

\begin{figure}
\centering
\includegraphics{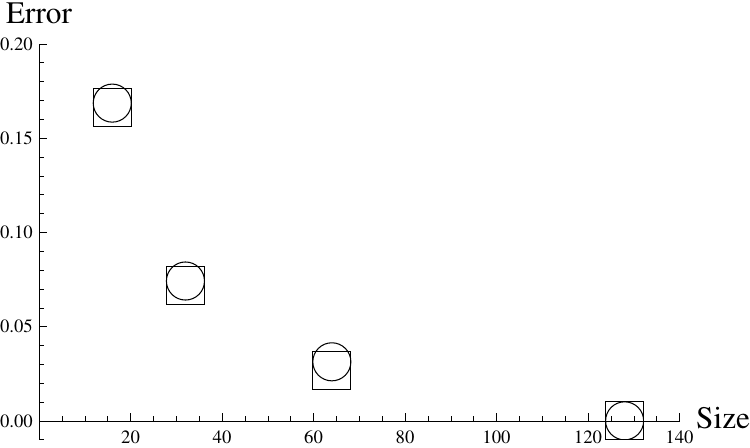}
\caption{\label{fig:LPConvergence} The errors with respect to the $128\times 128$ grids are presented for the $64\times 64$, $32\times 32$, and $16\times 16$ grids by the tensor product of the parametric model with $\zeta=4$ (square) and by the LBGK D2Q9 model (circle) for the simulation of the shear layer with periodic boundary conditions.}
\end{figure}

\subsection{Benchmark test showing enhanced stability and accuracy}
We demonstrate the enhanced stability of the parametric lattice Boltzmann model with $\zeta=4$ with a simulation of the shock tube. We use one thousand nodes ($1\leq \bar x \leq 1000$) for the linear shock tube. The initial condition is set by $C_L=\{\bar \rho_L, \bar u_L, \bar \theta_L \}=\{6,0,1\}$ for the left half space and $C_R=\{\bar \rho_R, \bar u_R, \bar \theta_R \}=\{1,0,1\}$ for the right where $\bar \rho$ is relative density with respect to a reference. Relative pressure $\bar p$ is obtained by the equation of state of ideal gas $\bar p=\bar \rho \bar \theta$. The physical properties of the extreme left and right are maintained by $C_L$ and $C_R$, respectively. Fig.~{\ref{fig:PLE6}} shows the results of flow velocity obtained by three different models; the parametric lattice Boltzmann model with $\zeta=4$, the LBGK model \cite{Qian1992} that is equivalent to the parametric model with $\zeta=3$, and the model obtained by an entropy function \cite{Ansumali2003}. The viscosity of the models is expressed by $\nu=(1/\omega-1/2)\sqrt {\theta_0}\Delta x/\sqrt{\zeta}$ so that we use $\omega=1$ for the LBGK and the entropic models because they share their discrete velocities, and $\omega=4\sqrt 3-6$ for the parametric model with $\zeta=4$ to match viscosity. We use the results after 362 iterations for the LBGK and the entropic models, and 418 iterations \footnote{The time mismatch is only about $0.0016(\approx 362 \times 2/\sqrt 3-418)$ iteration.} for the parametric model with $\zeta=4$. The LBGK model gives the unstable oscillating result (yellow solid line), while the parametric model with $\zeta=4$ (blue solid line) and the entropic model (red dashed line) provide the stable results. However, there is a disagreement on the velocity profile between the entropic model and the parametric model with $\zeta=4$. According to the analytic solution of the Euler equations with the Rankine-Hugoniot conditions, which is the same to the solution of the Navier-Stokes equations in the plateau regions of the shock profile, the parametric model with $\zeta=4$ gives accurate results as indicated on Fig.~{\ref{fig:PLE6}}. The reason is that the entropic model does not satisfy $\hat \mu_2=\mu_2$ in contrast to the LBGK model and the parametric model with $\zeta=4$ as listed in Table~\ref{tab:tableAccuracy}. Note that the moments $\hat \mu_2$ and $\hat \mu_3$ of the LBGK and the entropic models have the second-order accuracy in $u$, while the parametric model with $\zeta=4$ gives $\hat \mu_3=4\theta_0 u$. We have performed other simulations to investigate the effect of the moment errors of $\hat \mu_2$ and $\hat \mu_3$ of the models. The density, velocity, and temperature profiles of the LBGK model (white dashed), the parametric model with $\zeta=4$ (thick blue), and the entropic model (red dot-dashed) are shown in Fig.~\ref{fig:PLEsmall} for the initial density ratio $\bar \rho_L/\bar \rho_R=1.1$ in addition to the difference of density $\Delta \bar \rho$ for the parametric model (thick blue) and for the entropic model (red dot-dashed) with respect to the LBGK model. We observe that the differences are not easily observable for all the models. The maximum differences of density and velocity between the parametric model with $\zeta=4$ and the LBGK model are about $0.3$\%. Note that the difference between the LBGK model and the parametric model with $\zeta=4$ is much less than the difference between the LBGK and the entropic models when $\bar \rho_L/\bar \rho_R=4$. Instead of enhancing stability, the entropic model obtains serious damage in accuracy as in Fig.~{\ref{fig:PLE6}}. The deviation of the entropic model is noticeable when density ratio or flow velocity is relatively high. Especially in one-dimensional space, the viscosity $\nu=\frac{\zeta-1}{2}\times (1/\omega-1/2)\sqrt {\theta_0}\Delta x/\sqrt{\zeta}$ can be used for the three-velocities parametric model to exactly match the viscosity to that of the LBGK by considering $\hat \mu_3=\zeta \theta_0 u$. We present the simulation result of the shock tube that shows the difference between the parametric model with $\zeta=4$ and the LBGK is significantly reduced by this modification in Fig.~{\ref{fig:PLBMviscosityMatch}}.

We provide two-dimensional simulation of shear layers that generate vortices by the Kelvin-Helmholtz instability \cite{Minion, Dellar, Dellar2}. The initial condition is given by
\begin{equation}
u_x =
\left \{
\begin{array}{ll}
      u_0\tanh\left[l_0\left(y-\frac 1 4\right)\right] & \textrm{if } 0\leq y\leq \frac 1 2,\\
      u_0\tanh\left[l_0\left(\frac 3 4-y\right)\right] & \textrm{if } \frac 1 2<y \leq 1
    \end{array}
\right.
\end{equation}
and
\begin{equation}
u_y=u_0 \epsilon \sin\left[2\pi\left(x+\frac 1 4\right)\right] \textrm{ for } 0\leq x\leq 1 \nonumber
\end{equation}
where $l_0=80$, $\epsilon=0.05$ and $u_0=0.069$ for the domain of calculation $0\leq x \leq 1$ and $0\leq y \leq 1$ divided by $128$ by $128$ grids. The relaxation constants $\omega=1.99880$ and $1.99862$ are used for the LBGK D2Q9 model and the nine-velocities parametric model that is obtained by the tensor product of the three-velocities parametric model with $\zeta=4$, respectively. The relaxation constants are chosen to match viscosity. Fig.~\ref{fig:LPViscosityMatch2d} shows the simulation result obtained by the two isothermal models. The first two and the last two rows are obtained by the LBGK D2Q9 and the parametric models, respectively. The figures of the first and the third rows provide the velocity vectors (short orange arrows) with stream lines (long blue arrows) for time steps $500$, $1500$, and $1800$ (for the cases of the LBGK); and $577$, $1732$, and $2078$ (for the cases of the parametric model). The figures of the second and the fourth rows provide the vorticity for the same time steps with the contours of $\pm(0.01,0.02,0.05)$. The result of the parametric model is slightly unstable at time step $1732$ (equivalent to $1500$ for the LBGK) and we observe vortices at time step $2078$ (equivalent to $1800$ for the LBGK), however, that of the LBGK is already highly unstable at time step $1500$ and only noise is observable at the time step $1800$. Fig.~\ref{fig:lpCompare} shows the comparison of the velocity amplitude results of the shear layer simulation obtained by the tensor product of the parametric model with $\zeta=4$ (black thin line) and by the LBGK D2Q9 model (orange thick line) for the time steps from 500 (left subfigure of the first row) to 1500 (right subfigure of the second row) with intervals of 200 by the LBGK step. The contours indicates the values of 0.03, 0.07, and 0.08. The comparison shows the accuracy of the parametric model and the stability superior to the LBGK. Fig.~\ref{fig:LPConvergence} presents the errors with respect to the $128\times 128$ grids for the $64\times 64$, $32\times 32$, and $16\times 16$ grids by the tensor product of the parametric model with $\zeta=4$ (square) and by the LBGK D2Q9 model (circle) for the simulation of the shear layer with periodic boundary conditions. The errors are calculated for the velocity amplitude over the whole domain of calculation. The result shows the second order of convergence, which conforms to the proof of Junk and Yang \cite{Junk}.

\begin{figure}
\centering
\includegraphics{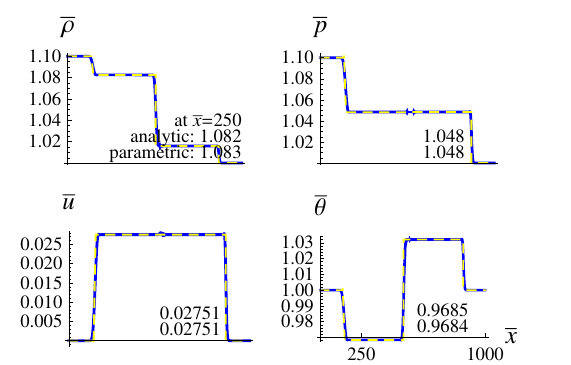}
\caption{\label{fig:PLBM5vshock} (Color online) The simulation result obtained by the model of five discrete velocities (blue solid line) are drawn with the analytical solution of the Riemann problem for the Euler equations (yellow dashed) for the purpose of a reference of the plateau values of the profiles. Note that the values at $\bar x=250$ are indicated on the figures.}
\end{figure}

\begin{figure}
\centering
\includegraphics{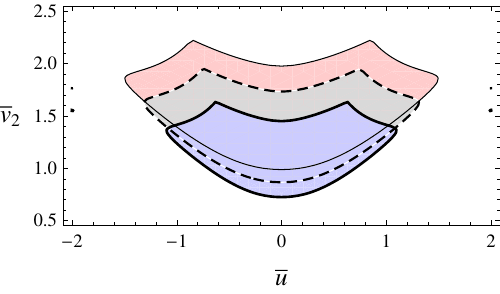}
\caption{\label{fig:5v} (Color online) The contour plot of the redistribution rule $r_i$ of the five discrete velocities is drawn when $v_4=2 v_2$. The blue region (thick solid boundary), the gray (dashed), and the red (thin solid) satisfy $r_i \geq 0$ for $\bar \theta =0.7$, $1$, and $1.3$, respectively.}
\end{figure}

\section{Analysis of the thermal models}

\begin{figure}
\centering
\includegraphics{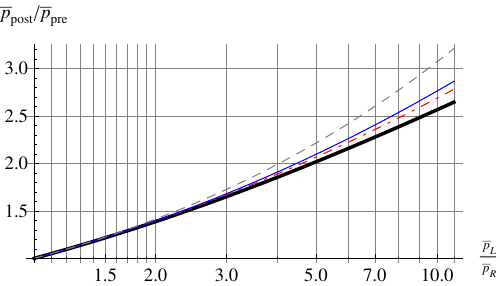}
\caption{\label{fig:Rankine} (Color online) The ratio between the post- and pre-shock pressures $\bar p_{post}/\bar p_{pre}$ with respect to the ratio between the high and the low pressures $\bar p_{L}/\bar p_{R}$ of an initial state is drawn by the solution of the Riemann problem of the shock tube for the Euler equations for the isothermal case (gray dashed), the cases of the one- (thin blue), the two- (red dot-dashed), and the three-dimensional spaces (thick black). The values of $\bar p_{post}/\bar p_{pre}$ with respect to specific values of $\bar p_{L}/\bar p_{R}$ are tabulated in Table~\ref{tab:tableRankine}.}
\end{figure}

The thermal compressible flow simulation with the five velocities model derived in Eq.~(\ref{eq:1d5v}) shows that the use of isothermal approximation must be done carefully even for the case of $\bar u \ll 1$. Fig.~\ref{fig:PLBM5vshock} shows the result obtained by the parametric model (thick blue) of five discrete velocities with $a=1.4$ and $b=2a$, which are selected by considering the ranges of $\bar u$, $\bar \theta$, and $\bar v_i$ that provide $r_i \geq 0$ as in Fig.~\ref{fig:5v}, and the analytical solution of the Riemann problem of the shock tube (yellow dashed) when $\bar \rho_L/\bar \rho_R=1.1$. The significant difference is observed in comparison to the isothermal models of three discrete velocities.  The flow velocity in the region of post-shock $\bar u_{post}$ and the shock speed $\bar u_{shock}$ obtained by the isothermal models are respectively over- and under-estimated by about $1.72$ times than the one-dimensional thermal case and by about $1.28$ times than the three-dimensional thermal case as well as the density profile having the well-known four steps instead of three steps, although the temperature fluctuation is about 3\%. This is due to the heat capacity ratio $\gamma$; the isothermal case $\gamma=1$ and the one-dimensional thermal case $\gamma=3$. According to the Rankine-Hugoniot conditions, we obtain $\bar u_{shock}$ and $\bar u_{post}$ by
$$
\bar u_{shock}=\sqrt{\frac{(\gamma+1)}{2}(\bar p_{post}/\bar p_{pre}-1)+\gamma}$$
and
$$
\bar u_{post}={ \left(\bar p_{post}/\bar p_{pre}-1\right)}/{\bar u_{shock}}$$
where $\bar p_{post}$ and $\bar p_{pre}$ are respectively pressures in post- and pre-shock regions. The ratio $\bar p_{post}/\bar p_{pre}$ with respect to  $\bar p_{L}/\bar p_{R}$ is provided in Fig.~\ref{fig:Rankine} and Table~\ref{tab:tableRankine} by the solution of the Riemann problem where $\bar p_{L}$ and $\bar p_{R}$ are respectively high and low pressures of initial states.

\begin{table*}
\caption{\label{tab:tableRankine}The values of the ratio between the post- and pre-shock presures $\bar p_{post}/\bar p_{pre}$ with respect to the ratio between the high and the low pressures $\bar p_{L}/\bar p_{R}$ of an initial state are tabulated for specific values of $\bar p_{L}/\bar p_{R}$ by the solution of the Riemann problem of the shock tube for the Euler equations for the specific heat ratios $\gamma=1$, $5/3$, $2$, and $3$ which are corresponding to isothermal, 3D thermal, 2D thermal, and 1D thermal cases.}
\centering
\begin{tabular}{c|cccc}

\hline
& isothermal&3D thermal&2D thermal&1D thermal\\
 $\bar p_{L}/\bar p_{R}$& ($\gamma =1$)&($\gamma =5/3$)&($\gamma =2$)&($\gamma =3$)\\ \hline
1.1&1.049&1.049&1.049&1.048\\ 
1.2&1.095&1.095&1.094&1.094\\ 
1.3&1.140&1.138&1.138&1.137\\ 
1.4&1.183&1.180&1.179&1.178\\ 
1.5&1.225&1.220&1.219&1.216\\ 
1.6&1.265&1.258&1.256&1.253\\ 
1.7&1.303&1.295&1.292&1.289\\ 
1.8&1.341&1.330&1.327&1.323\\ 
1.9&1.377&1.364&1.361&1.355\\ 
2&1.41&1.40&1.39&1.39\\
3&1.73&1.68&1.67&1.65\\
4&1.99&1.91&1.88&1.85\\
5&2.21&2.09&2.06&2.02\\
6&2.41&2.26&2.22&2.15\\
7&2.60&2.40&2.35&2.28\\
8&2.77&2.53&2.48&2.38\\
9&2.92&2.65&2.59&2.48\\
10&3.07&2.76&2.69&2.56\\
\hline 
\end{tabular}
\end{table*}

\begin{figure}
\centering
\includegraphics{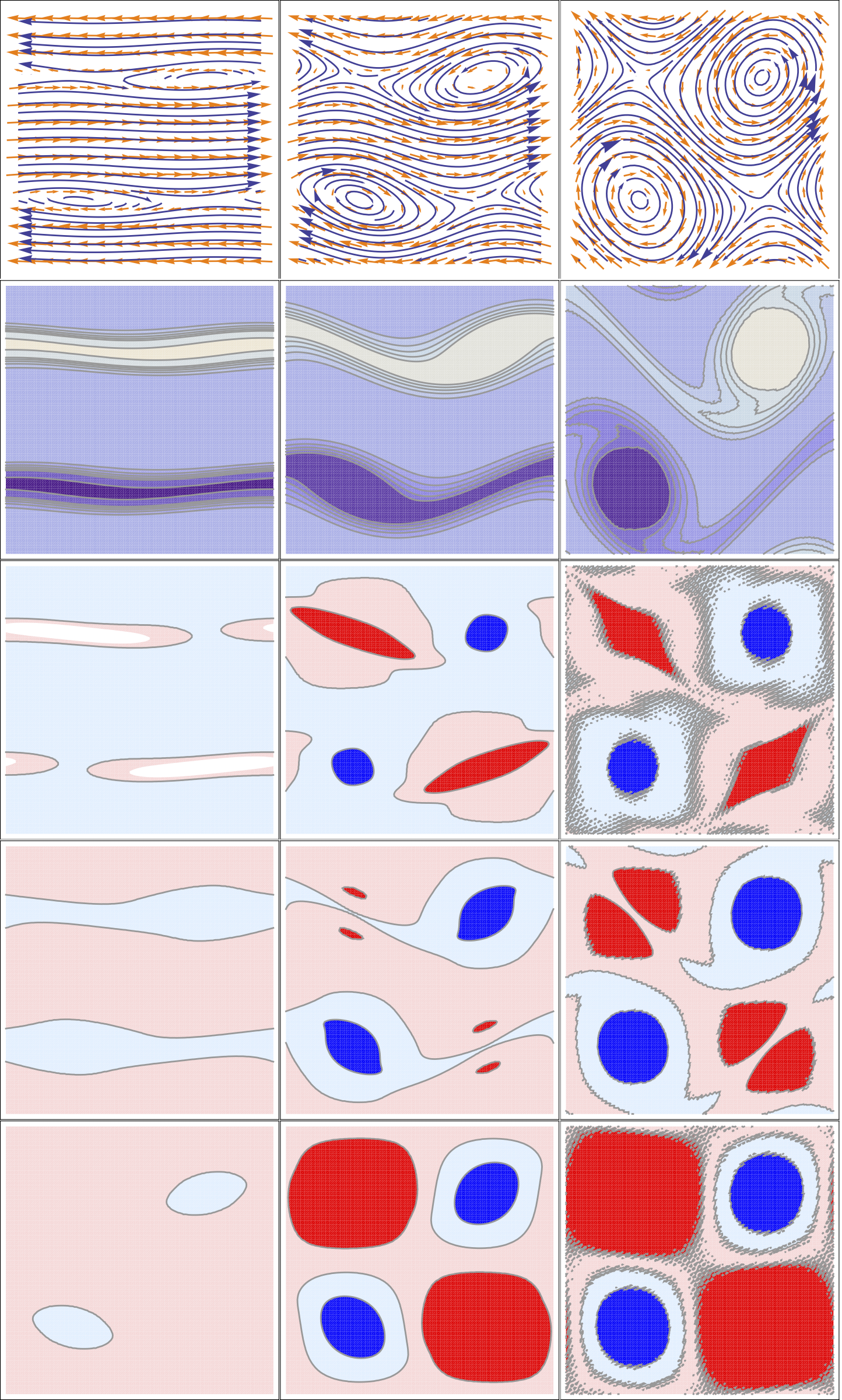}
\caption{\label{fig:PLBM25v} (Color online) The shear layer simulation in two-dimensional space is presented by the parametric 25-velocities model which is obtainable by the tensor product of the parametric five-velocities. The model recovers the fourth-order moment so that the accuracy is the level of the thermal Navier-Stokes equations. The figures of the first row show the velocity vectors (short orange arrows) with stream lines (long blue arrows) for time steps $924$, $2309$, and $3695$. The figures of the second, the third, the fourth, and the fifth rows provide the vorticity, the temperature, the density, and the pressure for the same time steps with the contours of $\pm(0.001,0.002,0.003,0.004,0.005)$, $(0.9995,1.0005,1.0015)$, $(0.9985,1.0000,1.0010)$, and $(0.9985,1.0000,1.0010)$, respectively.}
\end{figure}

We simulate the shear layer problem by the 25-velocities parametric model with $a=1.6$ which recovers the fourth-order moment and has the level of the accuracy of the thermal Navier-Stokes equations. In this simulation, the shear layers generate vortices by the Kelvin-Helmholtz instability \cite{Minion, Dellar, Dellar2}. The initial condition is given by
\begin{equation}
u_x =
\left \{
\begin{array}{ll}
      u_0\tanh\left[l_0\left(y-\frac 1 4\right)\right] & \textrm{if } 0\leq y\leq \frac 1 2,\\
      u_0\tanh\left[l_0\left(\frac 3 4-y\right)\right] & \textrm{if } \frac 1 2<y \leq 1
    \end{array}
\right.
\end{equation}
and
\begin{equation}
u_y=u_0 \epsilon \sin\left[2\pi\left(x+\frac 1 4\right)\right] \textrm{ for } 0\leq x\leq 1 \nonumber
\end{equation}
where $l_0=80$, $\epsilon=0.05$ and $u_0=0.069$ for the domain of calculation $0\leq x \leq 1$ and $0\leq y \leq 1$ divided by $128$ by $128$ grids. The value of $\omega=1.9$ is close to the upper limit for the given initial condition. In Fig.~\ref{fig:PLBM25v}, the first row shows the velocity vectors (short orange arrows) with stream lines (long blue arrows) for time steps $924$, $2309$, and $3695$. The figures of the second row provides the vorticity for the same time steps with the contours of $\pm(0.001,0.002,0.003,0.004,0.005)$. The figures of the third, the fourth, and the fifth rows provide the temperature, the density, and the pressure for the same time steps with the contours of $(0.9995,1.0005,1.0015)$, $(0.9985,1.0000,1.0010)$, and $(0.9985,1.0000,1.0010)$, respectively. We can observe that, in the areas where vortices occur, the temperature, the density, and the pressure are relatively lower than other areas. The numerical stability of the 25-velocities parametric model is demonstrated under the given initial condition in two-dimensional space. Note that one can use the 33-velocities on-lattice model \cite{Shim2013M} which has the level of accuracy of the thermal Navier-Stokes equations for lower viscosity and higher velocity flows.

\section{Conclusion}
In conclusion, we have presented parametric discretized equilibrium distributions of the lattice Boltzmann method. The ranges of flow velocity and temperature providing $r_i \geq 0$ vary with regulating discrete velocities as parameters. Relatively stable and accurate isothermal models are obtained. Thermal compressible flows are respectively simulated by only five on-lattice discrete velocities and 25 in one- and two-dimensional spaces in contrast to seven and sparse 33 or 37 velocities of conventional models so that the computational cost is reduced by about 30\%. The enhanced accuracy and the enhanced stability of the derived models have been tested and compared with existing models by the shock tube problem and by the shear layer problem in two-dimensional space. The equilibrium distributions upon asymmetric sets of discrete velocities are also introduced.

\section*{Appendix}

The redistribution rule $r_i$ corresponding to a set of discrete velocities $v_i$ for $i=1,2,\ldots,q$ is obtained by
\begin{equation}
\label{eq:constraints}
\sum_{i=1}^{q}v_i^{n}r_i=\int_{-\infty}^{\infty}v^{n}F(v) dv
\end{equation}
for $n=0,1,\ldots,n^*$ where $n^*$ is a desired order of accuracy,
$$F(v)=(2\pi\theta)^{(-d/2)}\exp[-\|v-u\|^2/(2\theta)],$$
$\theta=kT/m$, $k$ the Boltzmann constant, $T$ temperature, $m$ mass of a particle, $d$ dimension of space. In $d$-dimensional space with the Cartesian coordinate system, $v^n$ is defined by $\prod_{j=1}^{d} v_{x_j}^{n_j}$ for $n=\sum_{j=1}^{d} n_j$ with non-negative integers $n_j$ where $v_{x_j}$ is the $j$th coordinate component of $v$ for $j=1,\ldots,d$. In one-dimensional space for $n^*=q-1$, Eq.~(\ref{eq:constraints}) can be expressed by $R=V^{-1}M$ where
$$
V=\left [ \begin{array}{cccc}
    1 & 1 & \ldots & 1 \\
   v_1&v_2&\ldots&v_q\\
    \vdots & \vdots & \ddots& \vdots \\
    v_1^{q-1} & v_2^{q-1} & \ldots& v_q^{q-1}
  \end{array}
\right ],
R=\left [ \begin{array}{c}
r_1\\
r_2\\
\vdots\\
r_q 
\end{array} \right ],
M=\left [ \begin{array}{c} 
\mu_0\\
\mu_1\\
\vdots\\
\mu_{q-1} 
\end{array} \right ].
$$
By using the explicit expression of $V^{-1}$, we can express $r_i$ as
$$
r_i=\frac{\sum_{n=0}^{q-1} \left((-1)^n\mu_{q-1-n}\sum_{{1\leq j_1<\cdots<j_n \leq q-1 \textrm{  and  }
        {j_1\neq \cdots \neq j_n\neq i}}}  v_{j_1}\cdots v_{j_n}\right)}{\prod_{j\neq i}(v_i-v_j)}.
$$
\appendix
\section*{Acknowledgments}
This work was partially supported by the KIST Institutional Program.

\section*{References}
\bibliography{mybibfile}

\end{document}